\documentclass[12pt,preprint]{aastex631}

\begin{document}

\title{Comet C/2012 S1 (ISON) crossing the Jupiter orbit}

\author[0000-0003-4893-6150]{Margarita Safonova}
\affiliation{Indian Institute of Astrophysics, Bangalore 560 034, India}
\email{E-mail: margarita.safonova62@gmail.com}
\author{Noah Brosch}
\author{Shai Kaspi}
\affiliation{The Wise Observatory and the Raymond and Beverly Sackler School of Physics and
Astronomy, the Faculty of Exact Sciences, Tel Aviv University, Tel Aviv 69978, Israel}
\author{David Polishook}
\affiliation{Department of Earth, Atmospheric, and Planetary Sciences, Massachusetts Institute of Technology, Cambridge, MA 02139, USA} 
\author{R. Michael Rich}
\affiliation{Department of Physics and Astronomy, UCLA, Los Angeles, CA 90095-1547 USA}
\author{Firoza Sutaria}
\author{Jayant Murthy}
\affiliation{Indian Institute of Astrophysics, Bangalore 560 034, India}

\begin{abstract} 
We report results of intensive time-resolved imaging photometry and synoptic deep imaging of the comet C/2012 S1 (ISON) performed in February 2013. The data were obtained at the Wise Observatory in Israel (WO), at the Himalayan Chandra Telescope (HCT) in India, and at the Polaris Observatory Association in California, USA. During this period, the comet's heliocentric distance changed from 4.9 to 4.6 AU, just within the orbit of Jupiter. We analyze these early images in an attempt to determine the nuclear rotation period, assuming that at these relatively large heliocentric distances it would be possible to detect the photometric modulation of a rotating nucleus against an underdeveloped coma. Since this is not evident in our February 2013 data, with more than 400 independent photometric measurements analyzed, we can only set upper limits of 0.05 mag for periodic brightness modulations. We discuss (and discount) a possible brightening event (minor outburst) that occurred on $15-16$ February 2013.

We also present deep synoptic images of the comet, obtained by combining our exposures for each night, and analyze them. We find that during the period of our observations the comet exhibited a $\sim$$30^{\prime\prime}\simeq 60000$-km tail with no substructures visible and that this appearance did not change throughout our campaign.

The comet, as indicated by a single spectroscopic measurement obtained during this observation period, showed a dust coma reflecting the solar light. Our observations indicate that during February 2013, comet ISON was relatively quiet, with the dust coma presumably hiding any light modulation by a spinning nucleus.

\end{abstract}


\section{Introduction}

Long-period comets are believed to be composed of the primeval material that formed the Solar System, with most of their existence spent far from the Sun, in the Oort cloud region. Processes in that region, such as random disturbances from nearby stars or comet-comet collisions, may modify the orbits of Oort cloud comets and can direct some of them into the inner Solar System.
While residing in the Oort cloud, comets remain very cold, at temperatures close to that of the cosmic microwave background. However, once they start on their way towards the Sun, the heat influx onto them increases and various matter-release mechanisms come into play. The release of low sublimation temperature gases such as CO and N$_2$ may entrain with them grains of ices with frozen refractory materials. 

Comet C/2012 S1 (ISON) was discovered on 21 September 2012 (Novski et al. 2012) when it was $\sim$6.3 AU from the Sun and its total magnitude was 18.8. The discovery observations were quickly linked to pre-discovery images from 28 December 2011 by the Mount Lemmon Survey, and from 28 January 2012 by Pan-STARRS, allowing the determination of a preliminary orbit (Williams 2012). 
This showed that the comet may be on a parabolic or hyperbolic trajectory that will bring it to the perihelion on 28 November 2013 at a distance of only 0.012 AU (1,800,000 km) from the center of the Sun, slightly more distant than one and a half solar radii from the Sun's photosphere. The orbit showed also that on its way to perihelion the comet would pass relatively close to Mars, at 0.07 AU on October 1 2013. Sekhar \& Asher (2014) found a similarity in orbital parameters between comet ISON and comet C/1689 V1, speculating that they might have the same parent body.

While the close approach to Mars was not expected to affect the comet's appearance very much, it was clear that the close passage near the Sun would cause either the total disruption of the comet or, if it survives, at least bring about major changes in its physical parameters. Comet C1965 S1 (Ikeya-Seki) followed a trajectory that brought it within 450,000 km from the Sun on 21 October 1965; just before its perihelion passage, the comet's nucleus broke into (at least) three fragments and, after the perihelion, it appeared as one of the brightest comets of the last millennium.

Since major changes in the comet's appearance were expected post-perihelion, it was important to determine the basic characteristics of C/2012 S1 (ISON) as early as possible. One parameter that could be derived with the comet far from perihelion is its degree of activity by monitoring eventual changes in brightness and in the general appearance of the coma and the tail. Another is its nuclear spin, which requires the analysis of long photometric runs for periodic modulations. The importance of determining the spin of cometary nuclei has been stressed by Jewitt (1997). The time to determine the basic cometary properties and, in particular, its spin, is well before the comet enters the inner Solar System, preferably before the coma becomes dominant and masks the photometric modulation of a spinning nucleus.

For a long time it has been clear that nuclei of comets rotate. Rotation can be imposed on a cometary nucleus upon formation, possibly from a collision in the Oort or Kuiper belts, or from asymmetric outgassing activity once the comet enters the inner Solar System. The rotation can be detected directly as a periodic modulation of the light reflected from the comet when it is far from the Sun or when the nucleus is not active, since in other cases the brightness of the coma overwhelms the reflection from a small nucleus. The brightness modulation can originate from an irregular (elongated) shape of the nucleus, or from a non-uniform nuclear albedo distribution. Alternatively, sometimes it is possible to detect periodic structures (``waves'') in the cometary coma by judicious image processing, which can constrain, under certain conditions, the spin of the comet.

Jewitt (1997) gave in his Table~1 examples of cometary spin periods. These range from 6.1$\pm$0.05 hours for P/Wilson-Harr to 2.2 (7.2) days for P/Halley. We measured periods of 52 hours for 1P/Halley and 9.5 hours for 21P/Giacobini-Zinner (Leibowitz and Brosch 1985a,b) using photoelectric aperture photometry at the Wise Observatory, Israel. Samarasinha et al. (2004) listed in their Table~1 periods from 0$^d$.25 for 107P/Wilson-Harrington to 3$^d$.69 for comet Halley. Lamy et al. (2004) listed a range of rotational periods from 5$^h$ to 70$^h$. 

In principle, it is not necessary that the spin axis be aligned with the short axis of the nucleus, in which case the nucleus is considered to be in an excited state. Jewitt (1997) explained that the damping time-scale for the nucleus to leave this state is $\sim5 \times 10^6$ years for a 5-km nucleus. The damping time scale is inversely proportional to the square of the radius, thus larger nuclei damp their excited state faster and revert to having their spin axes aligned with their principal axis. Knight et al. (2013) measured the spin-down of comet 10P/Tempel 2 from 8.932$\pm$0.001 hours in 1988 to 8.950$\pm$0.002 hours in 2010. One possibility concerning comet ISON was that a significant spin change could take place following perihelion passage (e.g., Samarasinha \& Mueller 2013), thus the early monitoring campaign was justified.

The observations reported here were obtained when the comet was slightly closer than 5 AU, and are not the first photometric observations of C/2012 S1. The first systematic observations, except for sporadic observations by amateurs, were made by the space probe Deep Impact (DI, known in its extended mission as EPOXI) when the comet was slightly more distant than 5 AU. The 146 observations each 80-s long and exposed in a clear filter band, taken by the Medium-Resolution Imager of DI, span a 36-hour period on Jan. 17 and 18, 2013 (Agle \& Brown 2013). The data showed some variation in the comet brightness, but did not indicate a definite period.

The 5 AU heliocentric distance is approximately that of the ``frost line'' in the Solar System (Mumma et al. 2003), where hydrogen compounds in the solar nebula can condense into solid ice grains.
The EPOXI images show that although the comet was rather distant from the Sun, it was somewhat active already in January 2013, since it developed an elongated coma estimated to reach some 60,000-km from the nucleus. The photometry from the EPOXI measurements did not reveal periodic light modulations, but hinted at the possibility of a few small outbursts (Farnham, private communication).
Neither the HST (Li et al. 2013b) nor Spitzer (Lisse et al. 2013) detected periodic photometric variations in April 2013 or June 2013 respectively, but Lamy et al. (2014) reported a periodic $\sim10^h.4$ variation based on nine HST images obtained on November 1 2013.

We report here intensive observations of comet ISON obtained in February 2013 from Israel, India and the USA. Although we also did not detect periodic photometric variations or coma features that could reveal nuclear rotation, our observations and results should be reported to provide the community a full description of this comet's behavior. This paper is structured as follows: in Section~\ref{sec.obs} we detail the different observations and give a summary log. In Section~\ref{sec.proc} we detail the processing of the different types of data; this was done using mainly standard reduction programs. Section~\ref{sec.analysis} 
presents the results from the data analysis, and Section~\ref{sec.discuss} discusses these results in the context of the existing knowledge of cometary behavior compared to that observed from ISON. We present our conclusions in Section~\ref{sec.summary}.

\section{Observations and data reduction}
\label{sec.obs}

We imaged the comet with the C18 telescope -- 18-inch prime-focus reflector at the Wise Observatory (WO) in Israel (Brosch et al. 2008), with the 2.0-m Himalaya Chandra Telescope in India (HCT; Anupama 2001), and with a 28-inch prime-focus reflector in the US described in Rich et al. (2012) and called here C28US. The observing campaign was conducted in February 2013, with the most intensive observations taking place at the WO.

The largest fraction of the data analyzed here originates from the C18 telescope of the Wise Observatory, obtained with an SBIG STL-6303L thermoelectrically-cooled CCD without using a filter. On the C18, this CCD offers a 0$^{\circ}$.84$\times$1$^{\circ}$.25 field of view (FOV) with a scale of $1.4^{\prime\prime}$/pix. All C18 images were taken with open filter.

The HCT observations were performed with the Himalayan Faint Object Spectrograph and Camera (HFOSC) instrument that provides both imaging and spectroscopy. In imaging mode, the HFOSC offers a scale of $0.296^{\prime\prime}$/pix equivalent to a total FOV of 10$\times$10 arcmin$^2$ in its central 2k$\times$2k pixels, and images were obtained in the V, R and I Bessel bands. The readout noise, gain and readout time of the CCD are 4.87 e, 1.22 e/ADU and 90 sec, respectively.  

The C28US observations were collected with an SBIG STL-11K3 CCD through a luminance (L) filter, exposing for 180 sec. The 9 $\mu$m pixels of the CCD translate into a plate scale of $0.83^{\prime\prime}$/pix. Since the L filter is a very wide one, combining the V, R and I bandpasses while cutting off the blue and red ends of the band where terrestrial artificial and natural backgrounds are high, this implies that the C18 and the C28US used essentially the same bandpasses.

The WO C18 images were exposed for 120 sec using sidereal guiding. The guide star was not re-acquired during the observations, which resulted in images registered on the stars with the comet image steadily changing location on the chip. The HCT images were exposed slightly longer (upto 420 sec in January) and the V, R and I filters were used. A summary of all imaging observations is given in Table~\ref{t.images}, where we also list the typical seeing for each specific observation set. This was derived by measuring the FWHM of some five stars in images spanning the respective night, and choosing the median value as representative.
 
\begin{table}[h!]
\centering
\caption{Jan - Feb 2013 imaging observations of ISON}
\label{t.images}
\begin{footnotesize}
\begin{tabular}{|c|c|c|c|c|c|c|c|}
\hline
Date & UTC start & UTC end & $N$ & $\bar{r}$ (AU) & Observatory & FWHM ($^{\prime\prime}$) \\
\hline
22-23 Jan & 13:48 & 14:28& 6 & 5.0 & HCT & 2.2 \\
02-03 Feb & 21:17 & 01:13 & 92 & 4.9 & C18WO & 4.0 \\
03-04 Feb & 19:23 & 01:11 & 130 & 4.9 & C18WO & 3.5 \\
04-05 Feb & 19:19 & 21:11 & 36 & 4.9 & C18WO  & 3.5 \\
07-08 Feb & 19:21 & 00:32 & 101 & 4.8 & C18WO & 3.9 \\
08-09 Feb & 05:02 & 06:28 & 20  & 4.8 & C28US & 3.8 \\
08-09 Feb & 19:09 & 00:50 & 131 & 4.8 & C18WO & 3.4 \\
12-13 Feb & 05:57 & 07:56 & 28 & 4.8 & C28US & 3.4 \\
12-13 Feb & 19:02 & 19:37 & 13 & 4.8 & C18WO & 3.2 \\
15-16 Feb & 16:57 & 19:12 & 53 & 4.8 & C18WO & 3.5 \\
17-18 Feb & 21:06 & 21:25 & 8 & 4.7 & C18WO & 3.5 \\
18-19 Feb & 20:56 & 23:40 & 62 & 4.7 & C18WO & 3.6 \\
19-20 Feb & 16:10 & 17:10 & 4 & 4.7 & HCT & 2.0 \\
21-22 Feb & 13:35 & 13:40 & 1 & 4.6 & HCT & 2.4 \\
22-23 Feb & 18:02 & 19:10 & 6 & 4.6 & HCT & 2.2 \\
 \hline 
\end{tabular}\\
\vspace{0.2cm}
\noindent
Notes to Table~\ref{t.images}: $N$ is the number of useful images collected in the specific night. The parameter $\bar{r}$ is the average heliocentric distance of the comet for the observation night, taken from the JPL Horizons website at http://ssd.jpl.nasa.gov/redirect/horizons.html. The last column (FWHM) gives the typical seeing for the night.  
\end{footnotesize}
\end{table}

About one-third of the 691 images collected in our ISON observing program could not be used for the photometry analysis reported here. The few HCT images were obtained through filters, thus they were not compatible with the rest of the photometric data obtained with a much wider bandpass and were excluded from the period search, but were used to qualify the comet morphology. Some images from the C18 and C28US were rejected because the comet image merged with or was close to the image of a clearly visible background star. Other images, mainly from the HCT where the time allocation for this project was always near the full Moon, were affected by the enhanced background from moonlight. The analysis presented in the next section is based on the 423 deemed to be useful images from C18 and C28US telescopes.

To characterize the light from the comet we obtained a single 15-minute spectrum of the comet on 9 February 2013 at 22:26UT, using the Wise Observatory's FOSC instrument (Kaspi et al. 1995) on the 1.0-m telescope with a $10^{\prime}$-long  and $2^{\prime\prime}$-wide slit, aligned East to West, approximately in the direction of the apparent motion of the comet. The grism used for this observation, with 600 gr/mm, provides a dispersion of 3.1~\AA\, pixel$^{-1}$ and the spectrum covers the approximate range of 370 to 800 nm.

Along with the comet spectrum we obtained one spectrum of a solar analog star 55~Cnc. This is a G8V star, not exactly identical to the G2V spectral type of the Sun but sufficiently similar to it to be used for the indicative analysis shown below. Note that the HCT obtained a series of spectroscopic observations of comet ISON; this was used in Samarasihna et al. (2015). The full HCT spectroscopic and photometric observational data (Jan to Nov 2013) is available at NASA Planetary Data System archive (Safonova et al. 2019).

\section{Data reduction and analysis}
\label{sec.proc}

The imaging results were used to derive two research products presented here: the light variation of the central and brightest part of the comet as a function of time, from which it may be possible to derive a periodicity that could originate from the nuclear spin, and one daily deep ``synoptic'' image obtained by combining most of the images from each particular night.

To produce a light curve (LC), all the 423 selected images, where the comet was not blended with images of background stars, were measured. Only images collected with the Wise Observatory's C18 telescope, out of the 640, were free of background stars within the comet image and where the nucleus brightness could be measured.
To measure the brightness of the nucleus, we used a fixed 4-pixel ($\sim$$5^{\prime\prime}$) diameter aperture centered on the brightest part of the comet image. Since during the period of our intensive monitoring in February 2013 the Earth-comet distance changed by only $\sim$6\%, we can assume that the projected aperture size translates into $\sim$14,000-km at the comet.

\begin{figure}[t]
\centering{
\includegraphics[width=15cm]{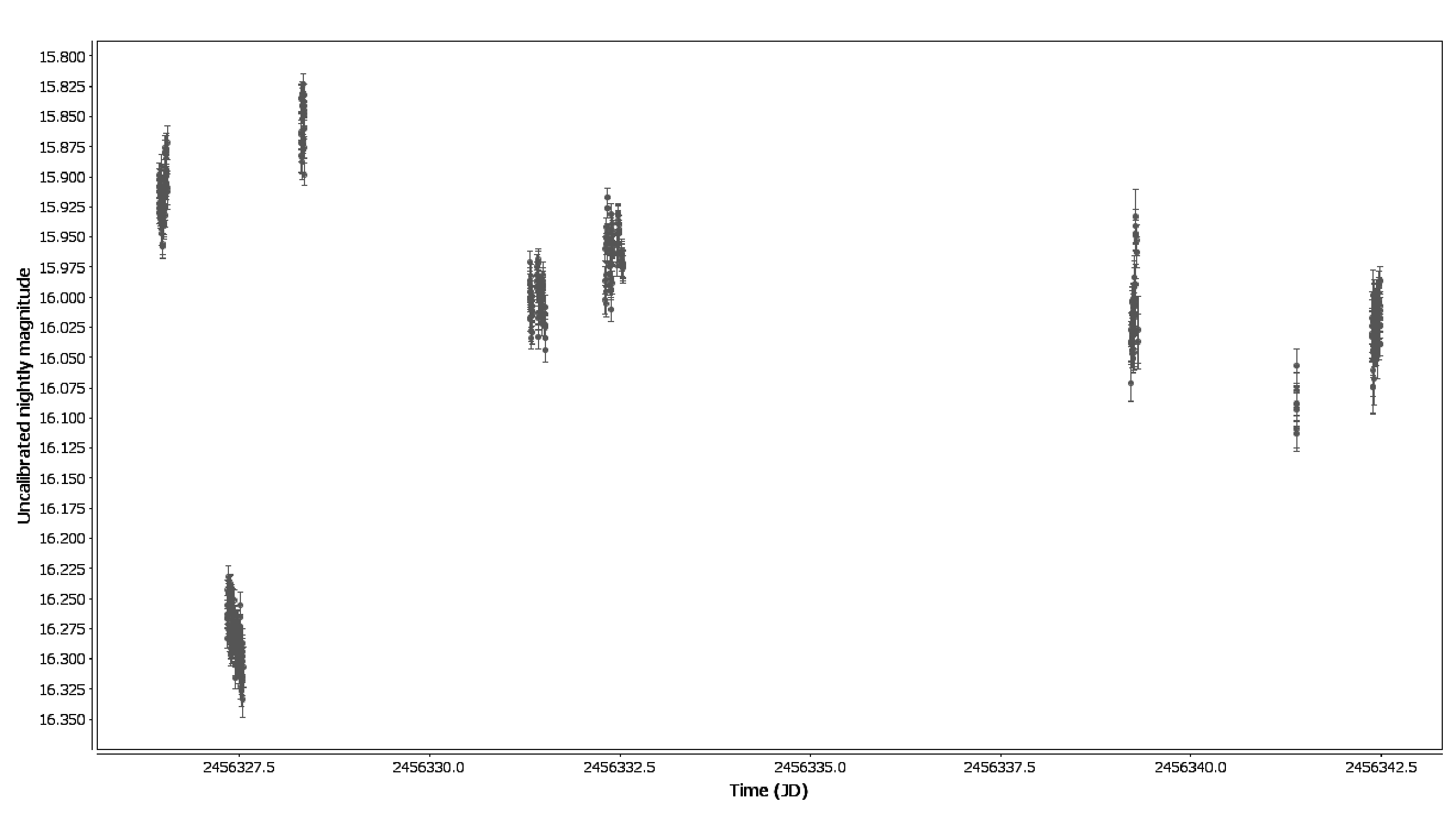}
\caption{Light curve of the near-nucleus region of comet ISON, measured through a four-pixel aperture.}
\label{fig:LC1}}
\end{figure}

We measured in the same way some tens of stellar images in the same frame, which served as local photometric standards. Comparing the brightness of the comet with that of an assembly of stars in the same image provides an uncalibrated magnitude for the comet while automatically compensating for changing atmospheric extinction and for transient and partial cloud coverage. The disadvantage of this method is that observations from different nights cannot be easily linked together, but will show an arbitrary shift in the zero level. The LC of comet ISON, as produced from the measurements described above, is shown in Figure~\ref{fig:LC1}.

In order to perform a period analysis, we subtracted from the comet nuclear magnitudes of each image the nightly average of these magnitudes (and added a constant $+15$ mag, to avoid negative values). This procedure yields a photometric stream averaging at 15 mag while retaining the possible inherent variability on time scales shorter
than 24 hours.

The light curve, after subtracting the nightly means, is shown in Figure~2. For clarity, no error bars are displayed in this plot, but they are, two ticks on the vertical axis. The LC shows typical nightly variations smaller than 0.1 mag (peak-to-peak) with the largest intra-night variation of $\sim$0.15 mag occurring during the night of 15-16 February 2013. The individual points carry photometric errors of order 0.01 mag. Error bars are shown in the expanded view for the night of 15-16 February 2013 (shown in Figure~\ref{fig:Feb15-16}). The displayed error bars for each point are derived from the photon statistics of the comet nuclear brightness and from that of the calibration stars.

\begin{figure}[h!]
\centering{
\includegraphics[width=10cm]{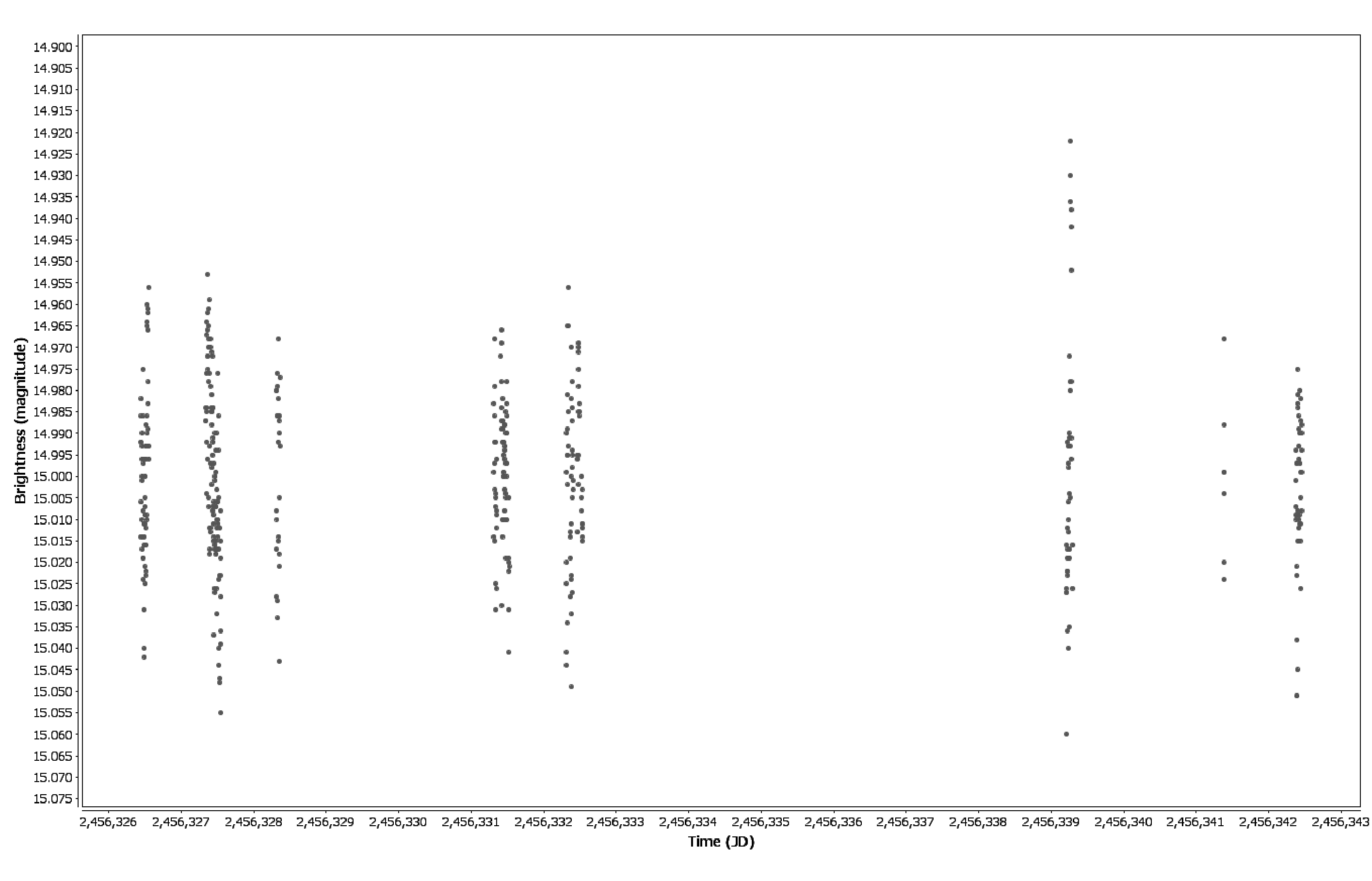}
\caption{Light curve of comet ISON after subtracting the mean nightly magnitude and adding a constant.} 
 }
\label{fig:LC2}
\end{figure}

\begin{figure}[h!]
\centering{
\includegraphics[width=10cm]{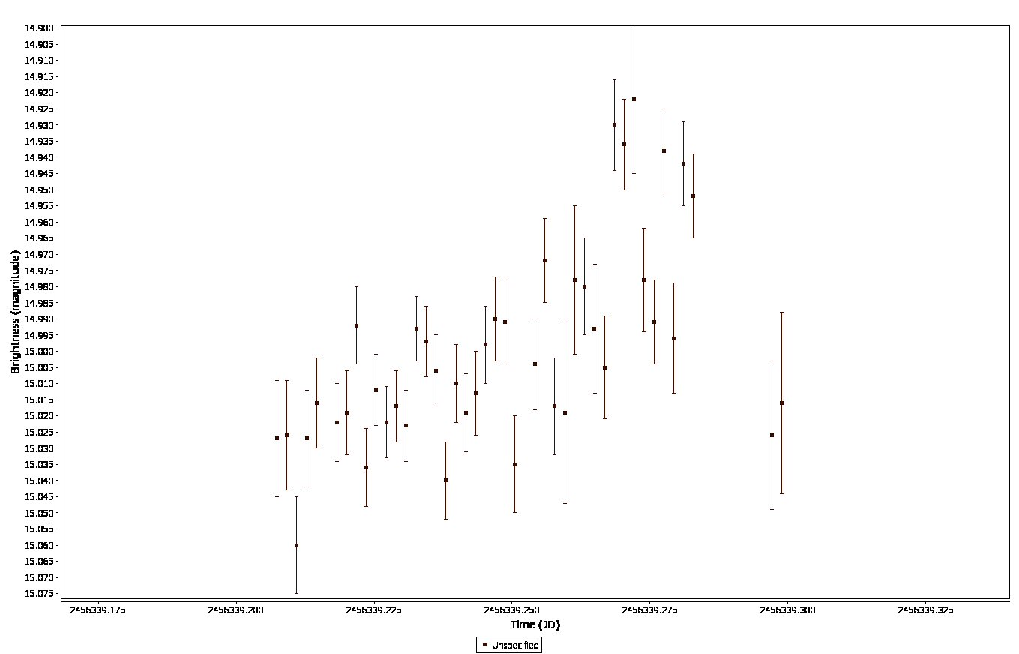} 
\caption{Light curve of comet ISON for the night 15-16 February, after subtracting the mean nightly magnitude and adding a constant.}
\label{fig:Feb15-16}
}
\end{figure}

\section{Analysis}
\label{sec.analysis}
\subsection{Time series and general photometry}

The means-subtracted LC was analyzed using the standard method used previously to investigate asteroid light curves, i.e. the Fourier analysis (Polishook \& Brosch 2009), as well as the AAVSO {\it v-star} package\footnote{http://www.aavso.org/vstar-overview}. Specifically, since our data are far from being equally spaced while covering 16 non-consecutive days, we searched for periodic variations when using {\it v-star} using the Data Compensated Discrete Fourier Transforms (DC-DFT: Ferraz-Mello 1981) and the Weighted Wavelet Z Transforms (WWZ: Foster 1996). The various attempts to fit periodic variations yielded upper limits of $\sim$0.05 mag amplitude for periods from 0.02 to 0.99 days. A specific search for periods around the 10$^h$.4 value reported by Lamy et al. (2014) was also negative.

\begin{figure}[h!]
\centering{
\includegraphics[width=10cm]{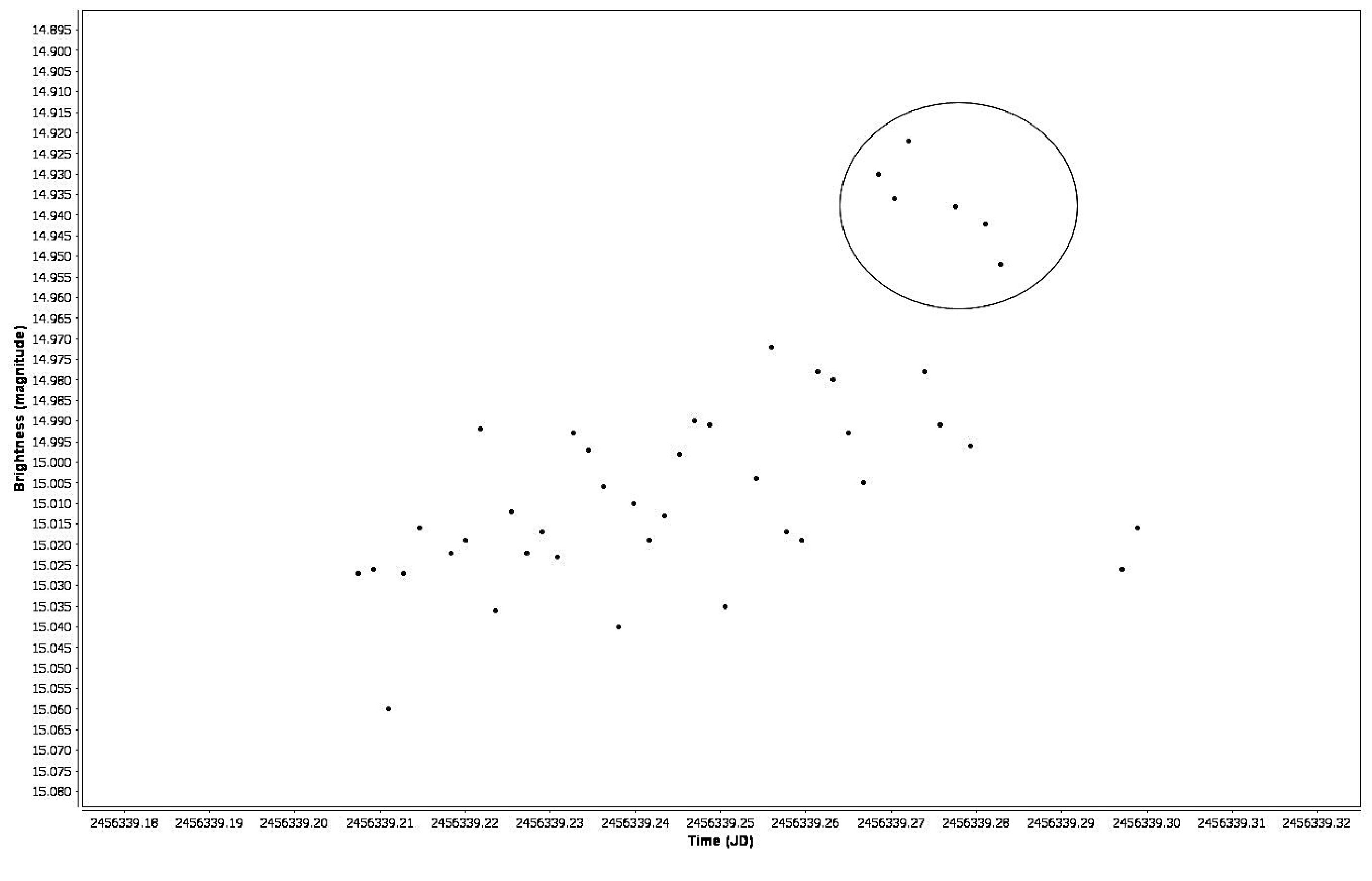}
\caption{Light curve for the February 15-16 night.}
\label{fig:Feb15-16-1}
}
\end{figure}

During our February campaign the means-subtracted LC showed a total variation of 0.15 mag with daily variations even smaller; typically of order 0.1 mag. The largest variation took place on the 15-16 February 2013 night, when a possible 0.05 mag brightening was recorded. The LC for this night is shown in Figure~\ref{fig:Feb15-16-1}. Close inspection of the measurements for this night indicates that only six points out of nine are above the general trend for this night, thus we cannot categorically classify the event as an outburst, although interpreting the brightness variation as a small outburst could be supported by the steady brightening of the comet in the $\sim$two hours of observation. The null assumption, that the brightness level was different from constant and not only randomly fluctuating, can be rejected with a $t-test$ that yields $t=-0.068$ with 42 degrees of freedom, indicating that this night's behaviour cannot confidently be interpreted as a small outburst.
 
\subsection{Spectrum and synoptic images} 

The comet spectrum was extracted from the long-slit WO image using a 16 pixels long aperture, corresponding to $27^{\prime\prime}$ on the sky. This is justified, given the size of the comet in the synoptic images. The spectrum of the solar analog was extracted with an 8-pixel wide aperture; this is $13^{\prime\prime}$ on the sky and implies that even the wide wings of the stellar image have been included in the spectrum. Both spectra were wavelength-calibrated using a He-Ar arc spectrum.

Given that the sky conditions during the observation were not photometric, and since the airmass difference between the comet and the solar analog star at the time of their observations was $\sim$0.2, we did not attempt to exactly compensate for the atmospheric extinction. Instead, we used a generic spectral extinction curve for Wise and the two airmasses of observation to calculate an approximate extinction correction. This is justified, since the color-dependent extinction coefficients at WO are of order $0.1-0.2$ (Vidal, Brosch \& Livio 1978) and, for an airmass difference of 0.2, the induced error would be of order a few 0.01\%.

To determine the general behaviour of the comet spectrum, we smoothed it along with  the spectrum of the solar analog with a 3-pixel ($\sim$10~\AA\,) running-mean filter. Since the blue ends of spectra obtained with the WO FOSC do not contain significant information due to the low FOSC lens transmission in the blue, both spectra were truncated to start from 4200~\AA\,. The smoothed and truncated spectrum of the solar analog was divided pixel-by-pixel by the spectrum of the comet. This procedure should, in principle, eliminate solar photospheric features and reveal cometary features that are not just the reflected solar light.

\begin{figure}[ht]
\centering{
  \includegraphics[width=12cm]{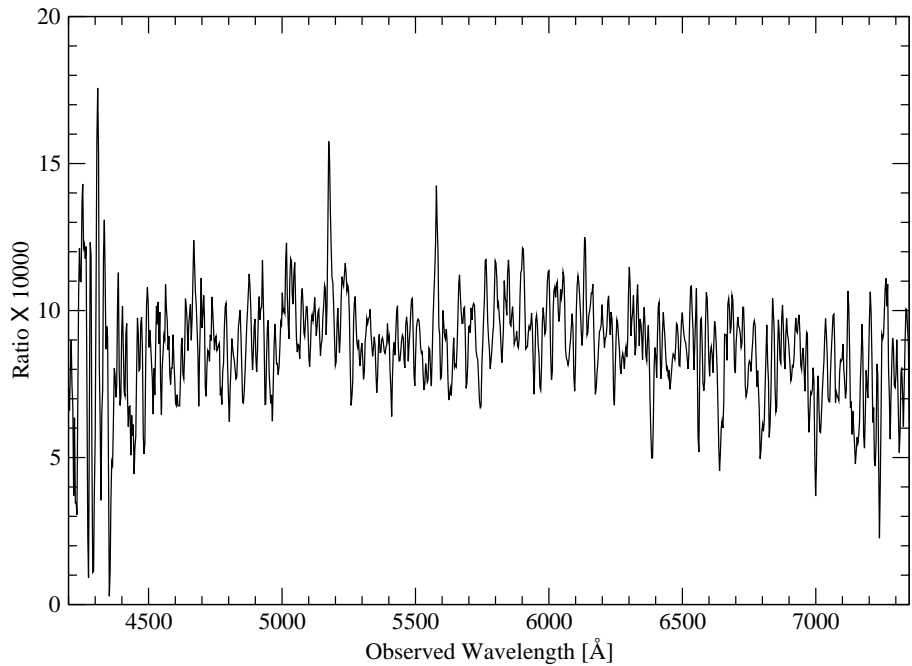}
  \caption{Relative reflectance of comet ISON: this is the result of the division of the comet spectrum by that of the solar analog.}
  }
  \label{fig:WO_spec-div}
\end{figure}

The result of the division of the solar analog by the comet spectrum, which is the relative spectral albedo, is shown in Figure~5. The spectral albedo appears approximately flat and does not show prominent features that could be interpreted as spectral line features. 
Since our comet spectrum from early February 2013 shows no prominent absorption or emission features, and the result of the division of the two spectra is essentially flat with wavelength, we conclude that the solar photospheric features had been properly eliminated by the division by the solar analog spectrum and that at the time of the observations the comet shone mostly by reflecting solar radiation.  

We mentioned earlier that the search for periodic light modulations yielded only the upper limits of $\sim$0.05 mag. This is not unexpected since, as will be detailed below, observations of the comet following our campaign showed that the nucleus was very small, less than one km in diameter. With such a small size, the nucleus would hardly cause a measurable signal on top of the coma brightness.
The synoptic nightly images were produced by selecting comet images where the comet was sufficiently distant from background stars, registering them on the comet's brightest part, and combining them with a median filter or with intensity averaging. While this produces reasonable comet images, the stars register as trails. Figure~6 presents a mosaic of the nightly images resulting from our program with the C18 telescope. Figure~7 shows the synoptic images from the HCT and C28US telescopes. Note that the first synoptic image from the HCT, which is the first ISON image obtained in this program, is the combination of six images, with each pair observed through one of the V, R or I filters.  

\begin{figure}[h!]
\centering{
  \includegraphics[width=5cm]{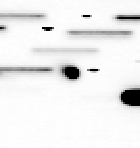}
  \includegraphics[width=5cm]{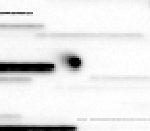}
  \includegraphics[width=5cm]{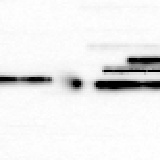}
  \includegraphics[width=5cm]{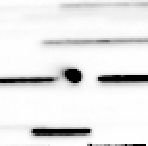}
  \includegraphics[width=5cm]{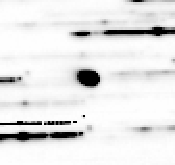}
  \includegraphics[width=5cm]{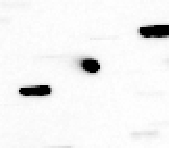}
  \includegraphics[width=5cm]{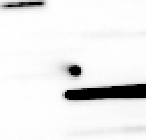}
  \includegraphics[width=5cm]{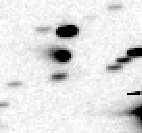}
  \includegraphics[width=5cm]{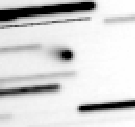}
\caption{Mosaic of daily synoptic images of comet ISON from the C18 telescope. The images were registered on the comet position, thus the stellar images appear trailed. The top row (left-to-right) shows the images from $2-3$, $3-4$, and $4-5$ Feb. The middle row in the same direction shows the images from $7-8$, $8-9$ and $12-13$ Feb. The bottom row shows the images from $15-16$, $17-18$, and $18-19$ Feb.}
}
\label{fig:synoptic}
\end{figure}

\begin{figure}[h!]
\centering{
\includegraphics[width=6cm]{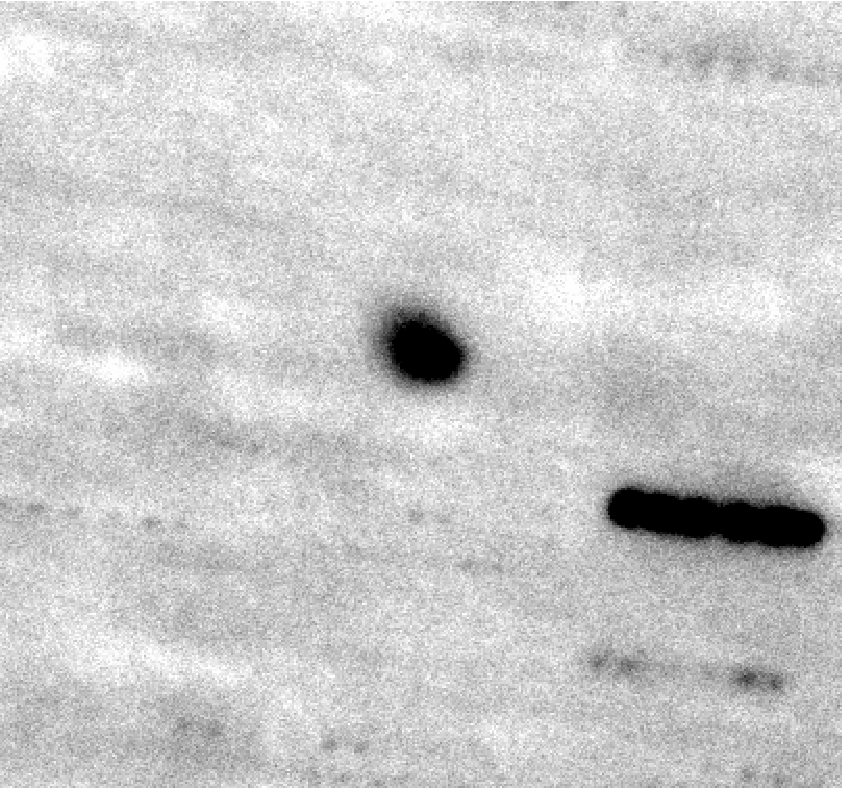}
\includegraphics[width=6cm]{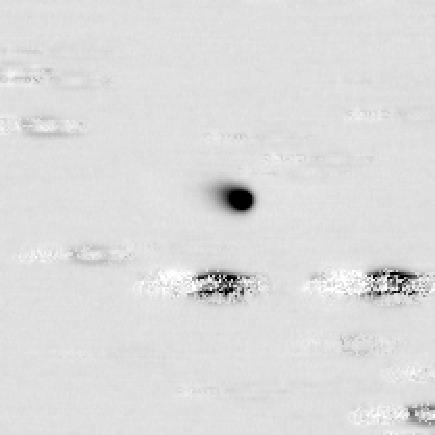}
\includegraphics[width=6cm]{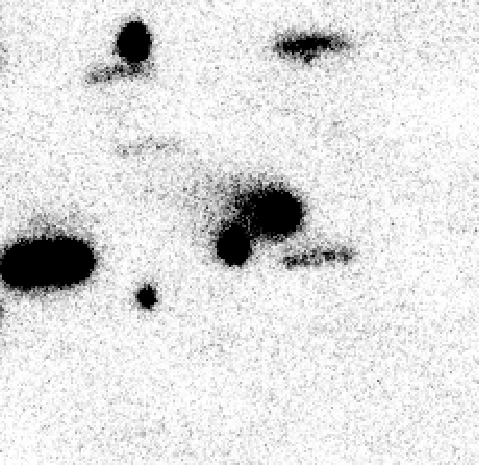}
\includegraphics[width=6cm]{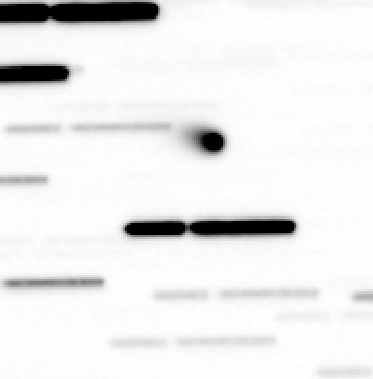}
\caption{Mosaic of daily synoptic images of comet ISON. HCT: 22 Jan (top left panel), 22 Feb (bottom left panel). C28US: 8 Feb (top right panel), 12 Feb (bottom right panel).}
  }
\label{fig:synoptic-other}
\end{figure}

Synoptic images offer a possibility to study the changing of the general aspect of the comet during the $\sim$two weeks of our intensive observing campaign. As the images presented in Figures~6 and 7 show, there are virtually no changes in the appearance of the comet. To demonstrate this, we plot in Figure~8 isophotes of the comet derived from our synoptic images from the night 18-19 February 2013 (C18) and from the night of 8-9 February from the C28US. The C18 image shows that the elongated coma, from the center of the brightest part, was $\sim$11 pixels=$15^{\prime\prime}$ long. 
The total length of the comet, from its leading edge close to the sky brightness level to the location where the tail merges with it, was at least $22^{\prime\prime}$. The corresponding values for the C28US, which produced deeper and better-sampled images, are a $25^{\prime\prime}$ tail and a $33^{\prime\prime}$ total length. The comet isophotes do not show twisting that might hint at rotation, nor do they show inner structures, possibly because of the limited angular resolution of our observations.

\begin{figure}[ht]
\centering{
  \includegraphics[width=7cm]{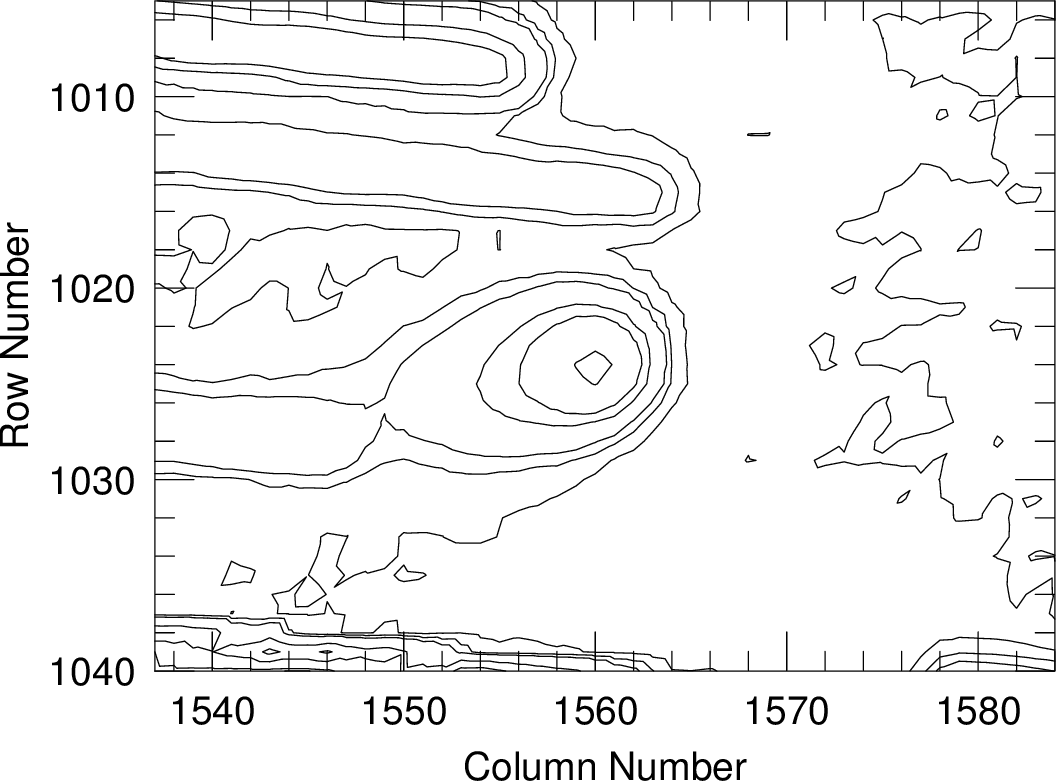}
  \includegraphics[width=7.3cm]{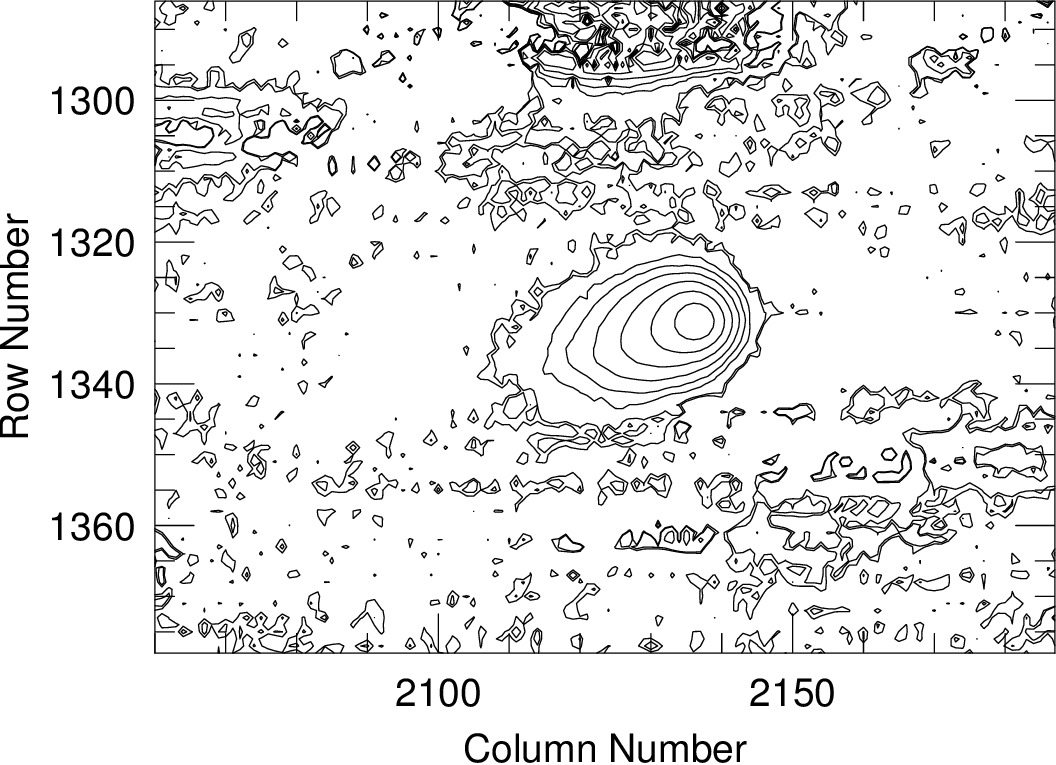}
\caption{Iso-intensity plots of the comet from the synoptic image of 18-19 February (C18: left panel) and from 8-9 February (C28US: right panel). The figures show that, within the angular resolution and exposure depth of this image, the comet's tail, or elongated coma, appears straight and no inner-coma structures are visible.}
  }
  \label{fig:isoph}
\end{figure}

\begin{figure}[ht]
\centering{
  \includegraphics[width=7cm]{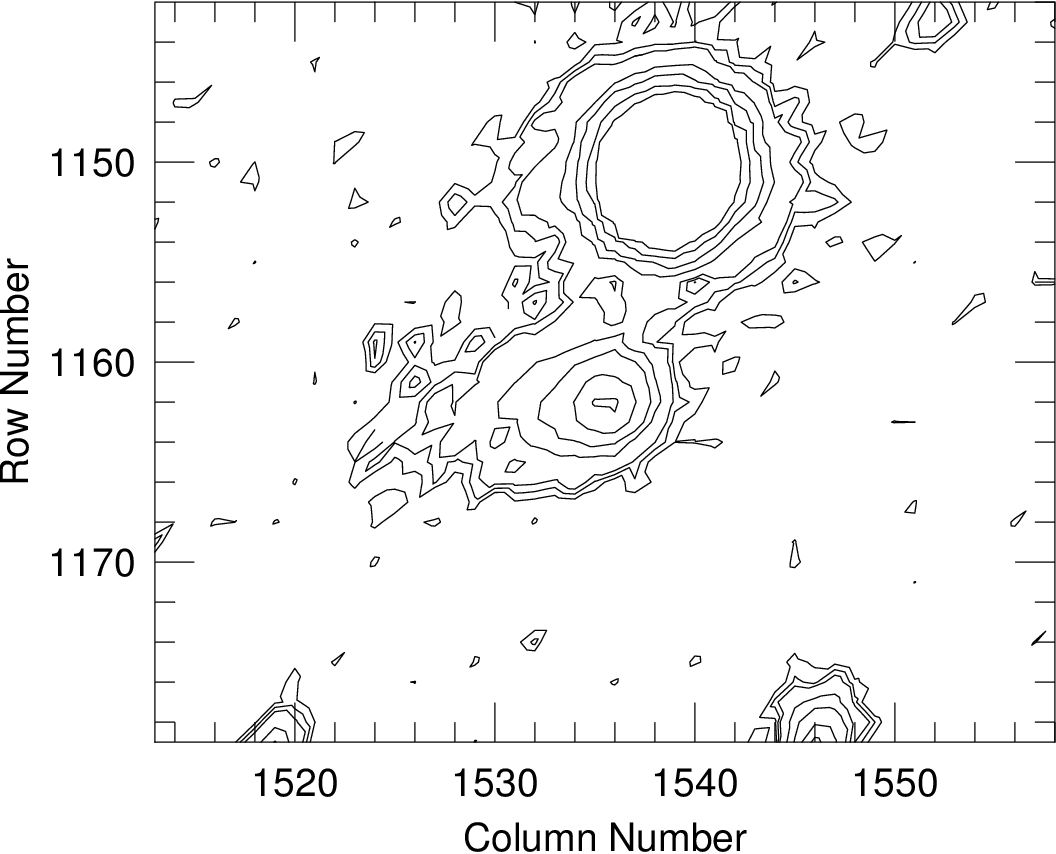}
  \includegraphics[width=7cm]{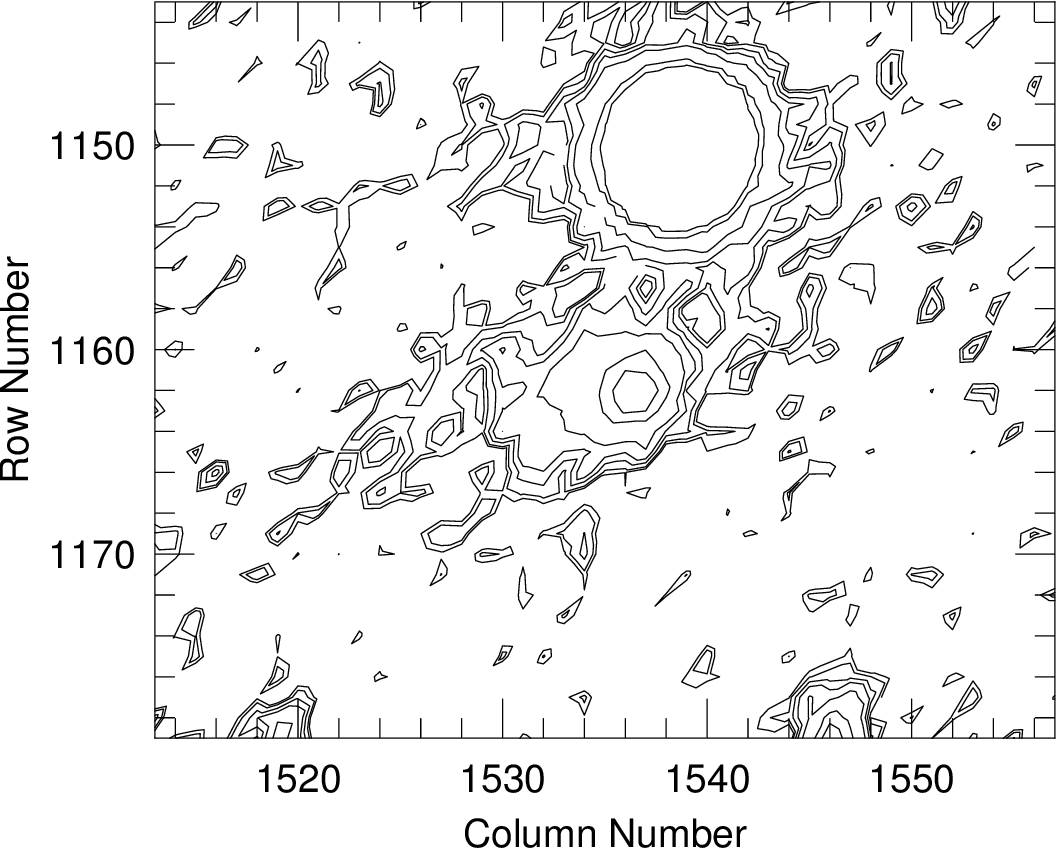}
\caption{Iso-intensity plots of the comet from two C18 adjacent images on 15-16 February. Left panel corresponds to the image where the inner part of the comet showed the highest brightness. Right panel shows similar isophotes for the image immediately following that with the peak brightness, i.e. some 2 minutes afterwards.}
  }
\label{fig:peak-isophot}
\end{figure}

We inspected the images related to the possible small outburst or brightening on $15-16$ February. Figure~9 shows the comet isophotes in two images adjacent in time: one corresponding to the peak of the light curve on that night and the other immediately following, corresponding to a low brightness point in the light curve. A comparison of the comet isophotes in the the two images plotted in Figure~9 indicates that the brightening might have been caused by the comet being very close to a relatively bright star. Although separated, so that the comet's nucleus could be measured, it is still possible that the wings of the bright nearby star  affected the measurement and an apparent peak was identified in the light curve.

\section {Discussion}
\label{sec.discuss}

\begin{figure}[t]
\centering
\includegraphics[width=8cm]{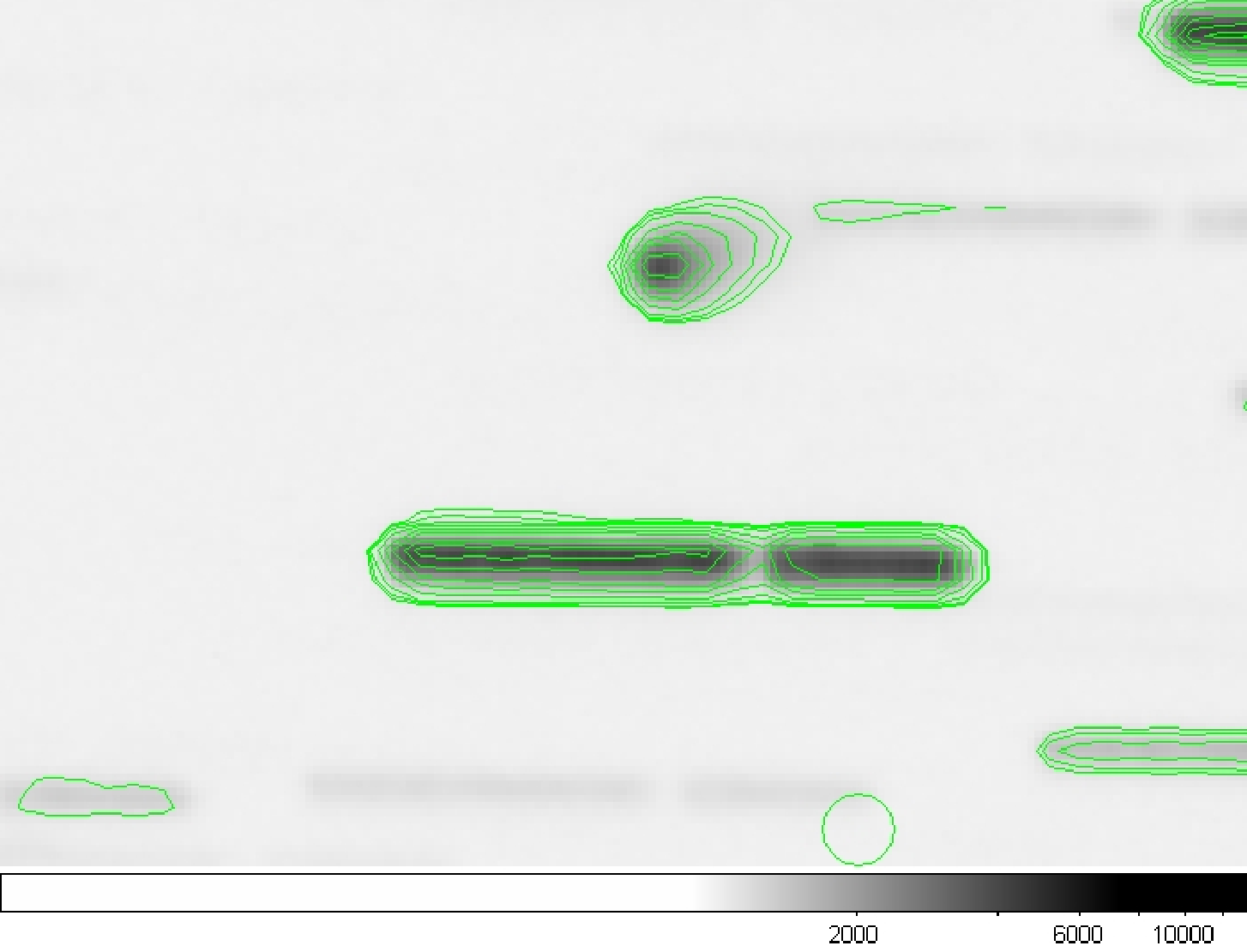} 
\caption{Deep image of comet ISON obtained by average-combining six HCT images from January 22, with iso-intensity contours. North is up, East is to the left, and sunward is to the left.}
\label{fig:ISON_HCT_contours}
\end{figure}

Our intensively-sampled light curve for February 2013 shows neither periodic modulation, with an upper limit of 0.05 mag, nor significant light outbursts. During this period the comet was essentially quiescent. This behaviour is confirmed by the EPOXI photometry (T.~Farnham, private communication) and by the light curve published by Scarmato (2014). An early report on the comet ISON's appearance is from the early EPOXI data, images collected every 15 minutes from January 17 03:40UT to January 18 16:20UT (NASA 2013) with the Medium Resolution Imager, a 12-cm aperture f/7.5 telescope with $2^{\prime\prime}$ pixels. During this period the comet showed a short anti-sunward tail or elongated coma, estimated to be at least 65000-km long. The plotted isophotes show a suspicious isophote twist at faint light levels. To investigate this, we repeated this analysis on our synoptic images and found this isophote twist in all our images. To demonstrate this, we plot in Figure~\ref{fig:ISON_HCT_contours} an image of the comet with over-plotted iso-intensity contours. Although affected by various imaging artifacts, this combined image also shows an apparent isophote twist of the faint contours. 

O'Rourke et al. (2013) detected the comet tail at 70 and 160 $\mu$m in March-April 2013, when ISON was $\sim$4.5 AU from the Sun, and claimed that the morphology of the 70~$\mu$m images indicates an excess of grains older than 60 days with slow ejection velocities ($\leq 10$ m/sec). Hines et al. (2014) observed comet ISON in early May 2013 with imaging polarimetry BY HST, using the WFC in the F606W band and with polarizing filters. From their observations, which showed a rotation of the polarization angle close to the nucleus, they conclude that the particles near the nucleus have different properties than those further away in the coma.

HST imaged the comet on April 10 2013 (Li et al. 2013b) and an enhanced version of this image shows a $1^{\prime\prime}.6$-long sunward-pointing wide jet that extends almost 4000-km from the nucleus, with an anti-sunward $30^{\prime\prime}$ tail and a total comet long axis of $\sim$20,000-km. No significant light variations were observed during the three-orbit HST observation, despite the superior angular resolution when compared with our images. The jet, reported also on earlier ground-based observations, remained unchanged from March to May 2013. Preliminary indications from the HST images were that the nucleus is less than 10 km wide.

In typical comets the coma and tail exhibit not only reflected solar radiation, but also emission from spectral lines and molecular bands. We investigated this aspect in comet ISON using spectroscopy when the comet was near the frost line in the Solar System and concluded that the light from the comet was essentially reflected solar light, i.e. only dust reflection.  

Upper limits to the nucleus size of order $1-1.2$ km were derived from the Mars Orbiter HiRISE images at end-September 2013 (Delamere et al. 2013).
Previous papers reported  an upper limit to the nucleus size of 2-km (Li et al. 2013a) with a lower limit of 0.8-km (Knight and Walsh 2013), and the presence of a sunward jet at 4.16 AU that did not change position in 19 hours (Li et al. 2013b).

Schleicher (2013) reported, from narrow-band photometry, the production rate of CN and of dust when the comet was at 4.57 AU from the Sun on 5 March 2013. Meech et al. (2013) proposed that the observed increased activity at 5.1 AU in January 2013 was due to the evaporation of a volatile layer. Note that Meech et al. (2009) asserted that at 5 to 6 AU heliocentric distance, the sublimation of H$_2$O ice can provide sufficient gas to entrain icy grains smaller than 0.1 $\mu$m off the nucleus, forming a coma. Their Table~1 gives examples of comets showing tails even at 30 AU.

In contrast with the spectacular pre-perihelion expectations, the behaviour of comet ISON following its perihelion passage was rather disappointing, with the comet disintegrating completely. The comet morphology immediately pre- and post-perihelion was described by Knight \& Battams (2014) from analyses of SOHO and STEREO images, benefiting from the capability of these platforms to observe the comet continuously while it was hidden by the Sun glare for ground-based observations. Knight \& Battams concluded that the small nucleus was destroyed just prior to perihelion leaving, if anything, a nucleus smaller than 10-m. The comet brightness was dominated by sunlight scattered off dust grains, as we showed above was the case already in February 2013. Sekanina (in Knight et al. 2013) analysed SOHO images very close to perihelion and concluded that the dust production terminated three hours prior to perihelion passage. Boehnhardt et al. (2013) reported on arclet-like features within the coma of comet ISON observed in mid-November 2013, which extended out of the nucleus a few 10$^3$-km on Nov. 14 and more than 10$^4$-km on Nov. 16, but were not present on Nov. 13. They suggested that the features were produced by sub-nuclei splitting off the nucleus as the comet approached perihelion. This matches the observed activity enhancement seen in OH (Crovisier, Colom, Biver and Bockelee-Morvan 2013), and in HCN (Biver, Agundez, Santos-Sanz, Crovisier, Bockelee-Morvan and Moreno 2013) near the same dates. These are presumably also connected with the optical outburst detected on Nov. 11-12 (Opitom et al. 2013a) that caused the comet to brighten by about one magnitude. Another optical outburst was detected on November 18-19 (Opitom et al. 2013b).  

Comet ISON is revealed by our measurements in February 2013 to have been a rather passive body, contrary to the expectations following its discovery but matching similar pre-perihelion observations. The various reports in the literature support a view that C/2012 S1 (ISON) was relatively quiet, forming an ion tail only close to perihelion in late-October 2013. Given that it is highly likely that ISON was a ``fresh'' comet originating from the Oort Cloud, this behavior is strange. The nucleus was likely very small, less than one km in diameter. It did outgas when approached perihelion, and it exhibited a significant coma and tail, but it did not become as spectacular as it was expected.

\section{Summary}
\label{sec.summary}

We reported here the results from an intensive imaging campaign on comet ISON during February 2013, using three different telescopes and collecting almost 700 independent images. About 60\% of these images were used to derive relative brightness measurements of the inner coma region of the comet. These measurements did not yield a credible periodic light modulation that could be interpreted as nuclear rotation to an upper limit of 0.05 mag, or a confident detection of a brightness outburst. 
Our images, including the deep synoptic ones derived daily for each observing station, had insufficient angular resolution to reveal coma details that could have hinted at rotation, jets, etc., but had shown that the comet's appearance was essentially stable in February 2013. Comet ISON was a small icy body fresh in from the Oort Cloud that disintegrated at perihelion without behaving in a spectacular manner.

\section{Acknowledgements}

We are grateful to the EPOXI team, in particular to Tony Farnham and Mike A'Hearn, for communicating their early results from the photometry of comet ISON. The observing time allocation committees of the telescopes used in this study are acknowledged for their generous time allocation. We thank the staff of IAO, Hanle and CREST, Hosakote, that made these observations possible. The facilities at IAO and CREST are operated by the Indian Institute of Astrophysics (IIA), Bangalore. DP is grateful to the AXA research fund for their generous postdoctoral fellowship.

\end{document}